\documentclass[twocolumn]{aastex63}
\usepackage{graphics,epsf}
\usepackage{amsmath}                
\usepackage{amsfonts}               
\usepackage{amssymb}                
\usepackage{epsfig}                 
\usepackage{appendix}
\usepackage{graphicx}
\usepackage{float}
\usepackage{color}
\usepackage{multirow}
\usepackage{colortbl}
\usepackage[para,online,flushleft]{threeparttable}

\newcommand{\km}{{~\rm km}}
\newcommand{\s}{{~\rm s}}

\newcommand{\erg}{{~\rm erg}}
\newcommand{\yr}{{~\rm yr}}

\newcommand{\eV}{{~\rm eV}}

\begin{document}

\title{Binary neutron star merger in common envelope jets supernovae}

\email{soker@physics.technion.ac.il}

\author{Noam Soker}
\affiliation{Department of Physics, Technion – Israel Institute of Technology, Haifa 3200003, Israel}
\affiliation{Guangdong Technion Israel Institute of Technology, Guangdong Province, Shantou 515069, China}

\begin{abstract}
I explore a triple-star scenario where a tight neutron star (NS) – NS binary system enters the envelope of a red supergiant (RSG) star and spirals-in towards its core. The two NSs accrete mass through accretion disks and launch jets that power a very luminous and long transient event, a common envelope jets supernova (CEJSN) event. Dynamical friction brings the two NSs to merge either in the RSG envelope or core. The total energy of the event, radiation and kinetic, is $\ga 10^{52} \erg$. The light curve stays luminous for months to years and a signal of gravitational waves might be detected. The ejecta contains freshly synthesized r-process elements not only from the NS-NS merger as in kilonova events, but possibly also from the pre-merger jets that the NSs launch inside the core, as in the r-process CEJSN scenario. This scenario shortens the time to NS-NS merger compared with that of kilonovae, and might somewhat ease the problem of the NS-NS r-process scenario to account for r-process nucleosynthesis in the early Universe. I estimate the ratio of NS-NS merger in CEJSN events to core collapse supernova (CCSN) events to be $\la 10^{-6} - 2 \times 10^{-5}$. However, because they are much more luminous I expect their detection fraction to that of CCSNe to be much larger than this number. This study calls for considering this and similar CEJSN scenarios in binary and in triple star systems when explaining peculiar and puzzling super luminous supernovae.
\end{abstract}


\section{Introduction} 
\label{sec:intro}

Common envelope jets supernovae (CEJSN) are transient events powered by jets that a black hole (BH) or a neutron star (NS) launch from within a red supergiant (RSG) star as they accrete mass from the envelope and then from the core of the RSG star  (e.g., \citealt{Gilkisetal2019, GrichenerSoker2019a, SokeretalGG2019, Schroderetal2020, GrichenerSoker2021}). 
Another type of explosion where an old NS accretes mass within a RSG is the explosion of a
massive Thorne–Zytkow object as \cite{Moriya2018} suggests. \cite{Moriya2018} studies a process where nuclear reactions in a massive Thorne–Zytkow object cannot support the envelope anymore and the NS accretes mass and launches jets. The main difference from the CEJSN scenario is that according to the CEJSN scenario  Thorne–Zytkow objects are unlikely to form (or if formed live for a very short time). Other differences is that in the CEJSN scenario the NS orbits the core of the RSG or within it, rather than being at its center, and that in the CEJSN scenario there is a much shorter time delay from the entrance of the NS to the RSG until explosion.

Two key processes allow the NS/BH to accrete mass at a high rate and launch the jets during this common envelope evolution (CEE), the formation of an accretion disk and neutrino cooling by the accreted mass. The density gradients in the  envelope and in the core leads to a non-axisymmetrical accretion flow where the NS/BH accretes more mass from the denser (inner to the orbit) side. The higher mass from one side results in a net specific angular momentum of the accreted gas that is sufficiently large to form an accretion disk around the compact NS/BH (e.g.,  \citealt{ArmitageLivio2000, Papishetal2015, SokerGilkis2018, LopezCamaraetal2019, LopezCamaraetal2020MN, Hilleletal2021}). 
Efficient neutrino-cooling by the dense and hot accreted mass for accretion rates of $\dot M_{\rm acc} \ga 10^{-3} M_\odot \yr^{-1}$ prevents the buildup of high-pressure around the mass-accreting NS/BH and allows the high mass-accretion rate to proceed \citep{HouckChevalier1991, Chevalier1993, Chevalier2012}. The jets themselves remove some more energy from the accreting object (e.g, \citealt{Shiberetal2016, Staffetal2016, Chamandyetal2018}). A black hole accretor also accretes some of the energy of the accreted mass (e.g., \citealt{Pophametal1999}). 
 
The energetic jets that the accretion disk around the NS/BH is very likely to launch  power the CEJSN and unbind large amounts of envelope mass. The removal of envelope mass by the jets can make the CEE efficiency parameter larger than unity, $\alpha_{\rm CE} > 1$, as some scenarios require (e.g. \citealt{Fragosetal2019, Zevinetal2021, Garciaetal2021}). As swell, the jets might be a site of r-process nucleosynthesis \citep{Papishetal2015, GrichenerSoker2019a, GrichenerSoker2019b} and for a BH accretor the jets might host the formation of very-high-energy ($\approx 10^{15} \eV$) neutrinos \citep{GrichenerSoker2021}.

If the NS/BH ejects the entire envelope before it enters the core the end product of the CEE itself is a NS/BH in a close orbit with the RSG core (e.g., \citealt{SokeretalGG2019}). Later the RSG core explodes and forms a NS/BH  (e.g., \citealt{SokeretalGG2019}). Else, the NS/BH enters the core and accretes at a very high rate as the NS/BH forces the destroyed core matter to form a massive accretion disk that most likely launches energetic jets (e.g., \citealt{GrichenerSoker2019a}). These jets power a CEJSN with explosions energies of up to  $\simeq {\rm several} \times 10^{52} \erg$. Because of the massive circumstellar matter (CSM) that the jets ejected at early phases of the CEE,  up to several solar masses, the ejecta-CSM collision leads to a long-lasting, months to years, bright transient (e.g., \citealt{SokeretalGG2019, Schroderetal2020}). 
CEJSNe might account for some puzzling super-energetic supernovae (SNe).  \cite{Thoneetal2011} suggested a CEJSN event where a NS merged with a helium star to explain the peculiar gamma ray burst (GRB) 101225A. \cite{SokerGilkis2018} suggested that SN~iPTF14hls (observations by  \citealt{Arcavietal2017}), and now also the similar SN~2020faa (observations by \citealt{Yangetal2021}), were CEJSN events, and \cite{SokeretalGG2019} suggested a version of the CEJSN scenario, the polar CEJSN scenario, to explain the fast-rising blue optical transient AT2018cow (e.g., \citealt{Prenticeetal2018, Marguttietal2019}). 
      
The large fraction of triple stellar systems among massive stars (e.g., \citealt{Sanaetal2014, MoeDiStefano2017}) motivate the exploration of CEJSN events in triples (e.g., \citealt{Soker20212CEJSNe}), in particular the entrance of a NS in a tight binary system into a RSG envelope. A tight binary system that enters a  CEE with a giant star might end in one of several outcomes as studies that did not consider NSs discuss (e.g., \citealt{SabachSoker2015, ComerfordIzzard2020, GlanzPerets2021}), including a merger process (e.g.,  \citealt{Hilleletal2017}). 
In an earlier paper \citep{Soker20212CEJSNe} I considered a tight binary system of a NS and a main sequence (MS) star that enters the envelope, and as a result of that the NS and the main sequence star merge. The NS accretes at a very high rate from the destroyed main sequence stellar matter, and launches energetic jets that power a transient event. Later the NS, or a BH remnant, might enter the core and launch energetic jets as it accretes from the core material. This double CEJSN event might be a long, months to years, luminous transient event with two bright peaks in its light curve.   
    
Here I study the CEE of a tight NS-NS binary system in the envelope and core of a RSG star. I describe this \textit{NS-NS merger in CEJSN} scenario in section \ref{sec:merger}, and estimate its rate in section \ref{sec:rate}. In section \ref{sec:r-process} I discuss implications to the r-process nucleosynthesis in the early Universe. In section \ref{sec:OtherCases} I list similar processes and place the NS-NS merger in CEJSN scenario in the context of other NS triple CEJSNe. I summarise in section \ref{sec:Summary}.

\section{The NS-NS merger in CEJSN scenario} 
\label{sec:merger}

\subsection{Evolutionary routes} 
\label{subsec:Routes}

\begin{figure*}
\hskip -0.90 cm
\includegraphics[trim=0.0cm 2.6cm 0.0cm 2.3cm ,clip, scale=0.90]{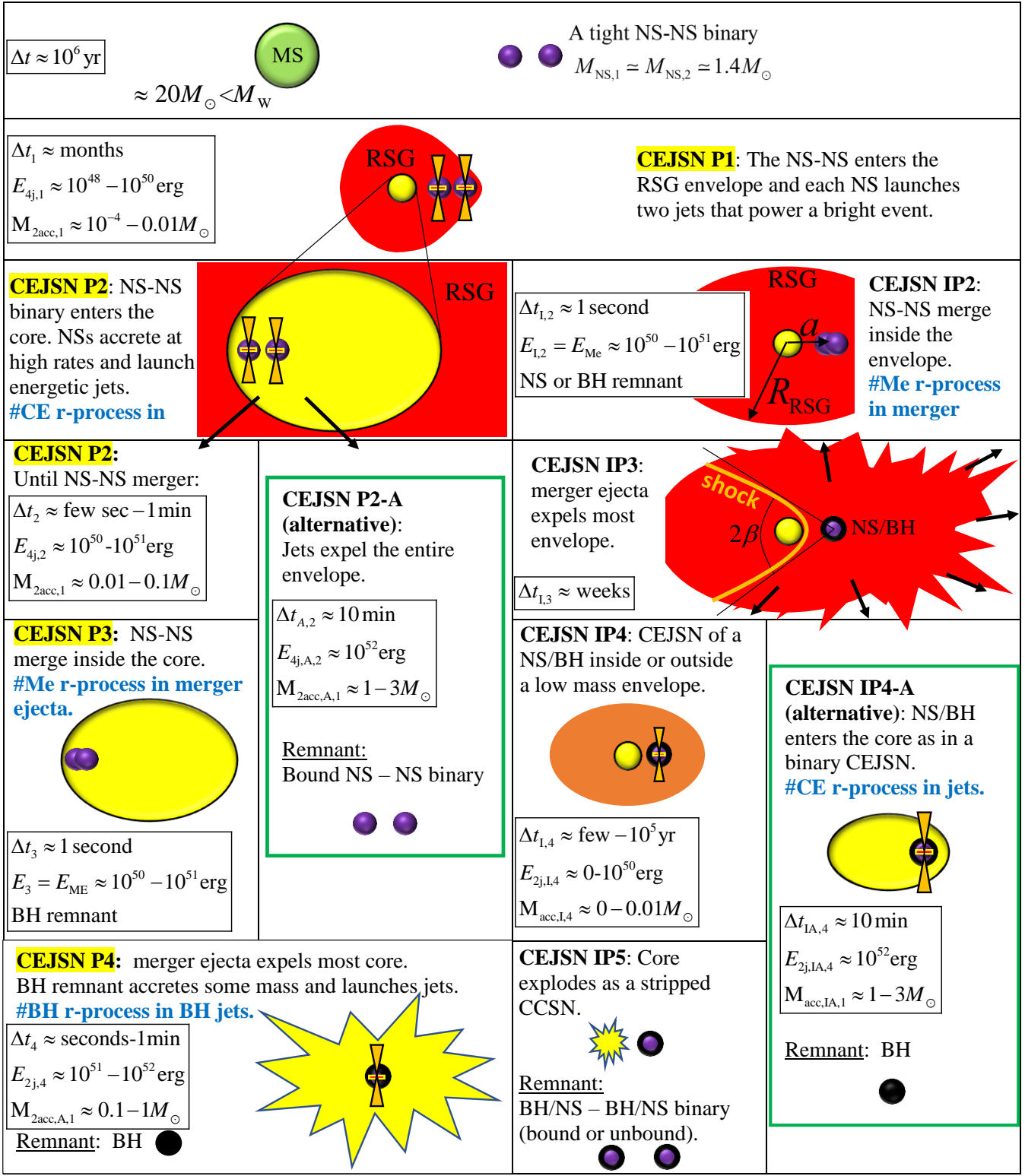}
\caption{A schematic diagram of the proposed NS-NS merger in CEJSN event. 
This figure serves to define some quantities that are not defined in the text. 
The first phase (P1) takes place when the tight NS-NS binary system enters the envelope. The second phase takes place either when the NS binary enters the core (P2) or when the two NSs merge inside the envelope (IP2). The case of the tight binary NS system in the core ends either with their merger (P3) and further mass accretion from the core material (P4) to form a BH remnant (left column), or, alternatively, the jets might explode the core and the system ends as two bound NSs. 
Merger inside the envelope releases a large enough energy to unbind most of the envelope (IP3), after which the core evolves (IP4) and explodes to form a NS/BH (IP5), or, alternatively, the NS/BH merger remnant from phase IP3 enters the core (IP4-A). The boxes list the duration, the total energy that the jets carry, and the mass that the NS/BHs accrete in the different phases. The r-process nucleosynthesis in the sites \#CE and \#BH takes place only if the core is oxygen rich (i.e., post-helium-core burning; see text). 
Abbreviation: BH: black hole; CCSN: core-collapse supernova;  MS: main sequence; NS: neutron star; RSG: Red supergiant. }
 \label{fig:SchematicScenario}
 \end{figure*}

I present the proposed scenario and its evolutionary routes in Fig. \ref{fig:SchematicScenario}. The figure itself defines the different phases of four evolutionary routes, three of which include NS-NS merger, and some quantities that I do not define in the text. As well, the figure presents the quantitative values of most evolutionary properties (with their large uncertainties) of the 4 evolutionary routes. The phases CEJSN P1 to CEJSN P4 (highlight by a yellow background in the figure) compose an extreme evolutionary route of the present study (section \ref{subsec:CoreMerger}) because it contains three potential r-process nucleosynthesis sites (section \ref{sec:r-process}). The last phase of the two routes where jets expel the core material before further evolution, i.e., no core explosion (phase IP4-A on the far right) or no NS-NS merger (phase P2-A in the second column), are inside greed rectangle.
In this figure I do not present other outcomes that are more similar to binary CEJSN events or CEJSN impostor (when the NS/BH does not enter the core) events. I mention them further in section \ref{sec:OtherCases}. These include the survival of the NS-NS binary outside the core and the core explosion, leaving three NSs (or a BH + NS-NS binary), bound or unbound. Another outcome is that one of the NS is ejected leaving one NS to orbit inside the RSG as in binary CEJSN. 

Before I continue with the proposed scenario I comment on the observational signatures that might distinguish the NS-NS merger in CEJSN scenario from the binary CEJSN event where only one NS enters the RSG envelope. The most obvious one is the occurrence of gravitational wave event because of the NS-NS merger during a bright and long-lasting explosion that mimic a CCSN. However, as far as electromagnetic signatures, once the tight binary NS enters the RSG envelope the differences might be too small to distinguish this case from a single NS that enters a RSG envelope and launches jets. A clear signature might be when the tight binary system starts to accrete mass from the RSG envelope before entering the RSG envelope, i.e., a pre-explosion signature. In that case there might be periodic photometric variations in the light curve, in the visible or more likely in X-ray, with a typical time period of hours to few days due to the orbital period of the tight NS binary system. By that time, though, the system is not as bright in the visible, but because the NS binary system is outside the envelope the system might be X-ray bright.  

The basic process that distinguishes the three focused CEJSN routes from others is the merging process of two NSs inside the envelope or inside the core (including cases when the destroyed core material forms a circumbinary accretion disk).  
Based on the observations of the NS merger (NSM) event GW170817 and simulations, studies (e.g., \citealt{Hajelaetal2021} and references therein) estimate the kinetic energy of the NS-NS merger ejecta as $E_{\rm Me} \approx 10^{51} \erg$ (e.g., \citealt{Trojaetal2020}) or somewhat below. I will take the energy range to be $E_{\rm Me} \approx 10^{50} - 10^{51} \erg$.
These studies conclude that  merger ejecta bulk velocities are $v_{\rm Me} \simeq 0.1-0.3 c$ (e.g., \citealt{Metzger2019}), and the merger ejecta mass is $M_{\rm Me} \simeq 0.06 M_\odot$ (the disk ejecta mass from the merger process is in the bulk range of $M_{\rm Me} \simeq 0.01-0.1 M_\odot$, e.g., \citealt{Metzger2019}). 

I will refer to the nucleosynthesis of r-process elements in NS-NS merger (kilonova;  e.g., \citealt{Metzger2017} and references therein) later on, but I do note here that \cite{Waxmanetal2018} argue that their claim for a relatively massive, $M_{\rm Me} \simeq 0.06 M_\odot$, ejecta with high velocities of $v_{\rm Me} \simeq 0.1 c$ to $>0.3 c$, and the low opacity that implies low Lanthanides fraction is in tension
with numerical simulations of NS mergers.
 
The exact geometry of the NS-NS merger ejecta is uncertain, but contains both equatorial and polar ejecta (e.g., \citealt{Metzger2019}). In the present flow structure both the equatorial and the polar ejecta collide with the dense RSG envelope and form a hot bubble around the merger site. For the purpose of this study I will take the hot bubble to be spherical. The merger remnant is a very massive NS or a BH (NS/BH). 

\subsection{NS-NS merger inside the envelope} 
\label{subsec:EnvelopeMerger}

Consider then a spherical explosion of $E_{\rm Me} \approx 10^{50} - 10^{51} \erg$ that takes place inside the RSG envelope (phases IP2 + IP3). Three dimensional hydrodynamical simulations of much weaker `explosions' due to the merger of two main sequence stars show that most of the explosion energy escapes (as kinetic energy of the expelled envelope gas) in the direction opposite to that of the giant core \citep{Hilleletal2017}. This is expected as the density towards the giant surface is much lower than towards the core. In other words, the RSG envelope gas that the ejecta expel to the opposite side of the RSG core takes most of the explosion energy. This gas escapes from the system at high velocities, i.e., much larger than the escape velocity. 

On the other hand, because most of the energy escapes the systems, the interaction of the merger ejecta with the RSG envelope gas within small angles $\la \beta$ to the direction of the core is more likely describes by momentum conservation (see schematic description of the interaction in phase IP3 of Fig. \ref{fig:SchematicScenario}). Presently I can only very crudely estimate the angle to be $\beta \approx 45^\circ$ based on the low energy `explosion' simulations of \cite{Hilleletal2017}. 
 
If the merger takes place at an orbital separation from the core that is much smaller than the RSG giant envelope, $a \ll R_{\rm RSG}$, the mass of the merger ejecta and the envelope mass that interact with each other under the momentum conservation assumption is 
$M_{\rm Me,\beta}\simeq 0.5 (1 - \cos \beta)M_{\rm Me}$ and 
$M_{\rm RSG,\beta}\simeq 0.5 (1 - \cos \beta)M_{\rm RSG,e}$, respectively,  where $M_{\rm RSG,e}$ is the total RSG envelope mass. For $\beta=45^\circ$ this fraction is $0.15$. 
The average velocity of the RSG envelope that the merger ejecta expel within small angles from the core direction is therefore, by momentum conservation, 
\begin{eqnarray}
\begin{aligned}
v_{\rm ex, \beta} & \approx v_{\rm Me} 
\frac {M_{\rm Me}}{M_{\rm RSG,e}} = 300
\left( \frac{ v_{\rm Me}} {0.1c} \right)
\\ & \times
\left( \frac{M_{\rm Me}} {0.1 M_\odot} \right)
\left( \frac{ M_{\rm RSG,e}}{10 M_\odot} \right)
\km \s^{-1} . 
\end{aligned}
\label{eq:VexBeta}
\end{eqnarray}

The escape velocity from a RSG of total mass $M_{\rm RSG} = 20 M_\odot$ at a radius of $1000 R_\odot$ is $v_{\rm esc} = 87 \km \s^{-1}$. This is much smaller than the scaled value in equation (\ref{eq:VexBeta}). 
However, there are other considerations. ($i$) The core deflects the shock wave from hitting the envelope gas behind it. Most of this gas that the shock does not hit directly might stay bound.   ($ii$) There is a velocity distribution and the outer parts of the expelled envelope move at velocities much large than $v_{\rm ex, \beta}$, while closer parts move at lower velocities.  ($iii$) For mergers of the same total energy but $v_{\rm Me} = 0.2c$ and ${M_{\rm Me}}=0.025 M_\odot$ equation (\ref{eq:VexBeta}) gives $v_{\rm ex, \beta} \approx 150 \km \s^{-1}$, which for a more massive RSG is not much larger than the escape velocity.  ($iv$) The merger energy can be smaller, down to $E_{\rm Me} \approx 10^{50} \erg$. 

The point is that some of the mass within small angles $\la \beta$ to the core direction, in particular behind the core, stays bound to the core-NS/BH binary system. 
I very crudely expect that in the case of a NS-NS merger in the envelope of a massive RSG star, $M_{\rm RSG,e} \ga 10 M_\odot$, the post-merger bound envelope mass is $M_{\rm PM,e} \approx 0.5-5 M_\odot$. 

Now the system can end in one of two ways. In one evolutionary route the NS/BH remnant does not enter the core (phase IP4) and the system evolves until the core explodes (phase IP5) leaving behind two NSs, bound or unbound. Because the NS/BH does not enter the core, this is sometimes termed a CEJSN impostor \citep{Gilkisetal2019}. This is the reason for `I' in the phase numbers.  In the second channel the NS/BH enters the core (phase IP4-A) and destroys it while launching energetic jets, as in binary CEJSN events. I refer to the r-process nucleosynthesis in section \ref{sec:r-process}. 
  
\subsection{NS-NS merger inside the core} 
\label{subsec:CoreMerger}
 
Consider next the evolutionary routes where a NS-NS tight binary system enters the core. One possibility is that the two NSs launch jets that expel the entire core material (phase P2-A) as in the binary CEJSN (like phase IP4-A). Observationally this is a CEJSN event that leaves behind a binary NS system instead of a single NS/BH. 
 
The extremest evolutionary route that I study here involves the merger of the two NS inside the core. If this process occurs before the jets of the NSs mange to eject most of the envelope, some core material is likely to stay bound. The consideration is as follows. The binding energy of the core, of mass $M_{\rm core}$ and radius $R_{\rm core}$, to the NS-NS tight binary system when the binary system enters the core is
\begin{eqnarray}
\begin{aligned}
E_{\rm b, core} \simeq 10^{50}
\frac {M_{\rm 2NS}}{2.8 M_\odot} 
\frac {M_{\rm core}}{2.5 M_\odot} 
\left( \frac{R_{\rm core}}{0.1R_\odot} \right)^{-1}
\erg,
\end{aligned}
\label{eq:Ebcore}
\end{eqnarray}
where $M_{\rm 2NS}$ is the combined mass of the two NSs.
Typical values for an RSG that is a descendant of a main sequence star of mass $M_{\rm ZAMS} = 30 M_\odot$  (model from \citealt{GrichenerSoker2019a}) are 
$(M_{\rm core},R_{\rm core})=(5 M_\odot, 1.2 R_\odot)$ when the star expands the first time and has a helium-rich core, and $(M_{\rm core},R_{\rm core})=(2.5 M_\odot, 0.09 R_\odot)$ when the star expands the second times with large quantities of oxygen in the core. In the case of a tight NS-NS binary system that merges inside a helium-rich core the energy that the merger process liberates is $E_{\rm Me} \gg E_{\rm b, core} \approx {\rm few} \times 10^{49} \erg$. Therefore, most likely the merger process ejects the entire core, as even the self-binding energy of the core will not be high enough to prevent complete ejection.  

In the case of a NS-NS binary that merges inside a more evolved core some core mass might stay bound. The merger will take place off-center, and as I discussed in section \ref{subsec:EnvelopeMerger} for merger in the envelope, material near the center might stay bound to the merger remnant. Because the NSs accreted mass before merger, the merger remnant is likely to be a BH, that now accretes mass from the left-over core mass (phase P4).

\section{Event rates} 
\label{sec:rate}

The merger of a NS-NS binary inside an RSG star is an early merger with respect to the time it would take the NS-NS binary to merge by the process of gravitational waves emission alone. It occurs at $\approx 10^7 \yr$ after star formation, i.e., the time it takes the wider companion to evolve to a RSG star. I will refer to the CEE induced merger as an early NS-NS merger process (EMe). The time from star formation to merger is the evolutionary time of the wide tertiary star to become a RSG. 

In their population synthesis study \cite{Schroderetal2020} find that the frequency of NS/BH single stars that enter a RSG envelope is $\simeq 1.5 \times 10^{-4} M^{-1}_\odot$, i.e., rate per solar mass of star formation, where in about half of the cases a NS enters the envelope. This rate corresponds to a fraction of $f_{\rm CEE} \simeq 0.026$ of all CCSNe (\citealt{Chevalier2012} estimated this ratio to be $\simeq 0.01$). 
Since the CCSN rate is the rate of NS/BH formation, a fraction of $f_{\rm CEE}$ of single NS/BH enters the envelope of RSG stars. If I assume, an assumption that needs confirmation by future simulations of triple-star evolution, that the same ratio holds as an upper limit (as not all NS-NS binaries that enter RSG stars merge) for the NS-NS merger in CEJSN scenario, then CEJSN might cause early NS-NS mergers to a fraction of about  
\begin{equation}
    f_{\rm EMe/Me} \equiv \frac{\rm Very~early~mergers}{\rm NS-NS~mergers} \la f_{\rm CEE} \approx 0.026
\label{eq:EarlyToNSNS}    
\end{equation}
of all NS-NS relevant binaries that will suffer merger. 

To find the ratio of the NS-NS merger in CEJSN (very-early mergers) to all CCSNe, I start with the fraction of NS-NS merger out of all CCSNe, $f_{\rm Me/CCSN}$, and multiply it by $f_{\rm EMe/Me}$. From the results of \cite{Safarzadehetal2019a} that give the rate of NS-NS merger per unit stellar mass formation, the fraction of NS-NS mergers to CCSNe is $f_{\rm Me/CCSN} \approx 3 \times 10^{-5} - 8 \times 10^{-4}$. 
The fraction of NS-NS merger in CEJSN to CCSN is therefore 
\begin{eqnarray}
\begin{aligned}
 &   f_{\rm EMe/CCSN}  \equiv  \frac{\rm Very~early~mergers}{\rm CCSNe} 
    \\ & \simeq  f_{\rm EMe/Me} \times f_{\rm Me/CCSN} \la 10^{-6} - 2 \times 10^{-5}.
\end{aligned}
\label{eq:eq:EarlyToCCSN}
\end{eqnarray}

The fraction of NS-NS binaries that enter the core (left two columns of Fig. \ref{fig:SchematicScenario}) is lower even. \cite{Schroderetal2020} find that a fraction of $f_{\rm core}=0.22$ of single NS/BH that enter the RSG envelope also enters the core of the RSG star. I take the same ratio to hold for binaries. Like \cite{Soker20212CEJSNe} I take that a fraction of $f_{\rm CO} \simeq 0.3-0.5$ of these cases occurs for post-helium-burning RSG stars. The three r-process nucleosynthesis sites take place one after the other only for oxygen-rich cores (section \ref{sec:r-process}).  
Over all, the fraction of the evolutionary route of three r-process sites (phases P1-P4 for oxygen-rich core) relative to all CCSNe is  
\begin{eqnarray}
\begin{aligned}
 &   f_{\rm 3r/CCSN}  \equiv  \frac{\rm Three~r-process~sites}{\rm CCSNe} 
 \\ & \simeq  f_{\rm EMe/CCSN} \times f_{\rm core} \times  f_{\rm CO}
    \la 5 \times 10^{-8} - 2 \times 10^{-6}.
\end{aligned}
\label{eq:eq:3rToCCSN}
\end{eqnarray}

\section{Implications to r-process nucleosynthesis} 
\label{sec:r-process}

The extremest evolutionary route that I study here, phases P1-P4 on the left column of Fig. \ref{fig:SchematicScenario}, involves three sites of r-process nucleosynthesis in the case that the process takes place when the core is post-core-helium burning (i.e., oxygen-rich core). 

\textit{\#CE The CEJSN r-process scenario.} The first r-process nucleosynthesis in the NS-NS merger in CEJSN scenario takes place when the two NSs accretes mass from the dense core (hence the demand for a post-helium-burning core rather than a helium-rich core; \citealt{GrichenerSoker2019a}). This is a process like a single NS that enters such a core in the CEJSN r-process scenario.

\textit{\#Me The kilonova (NS-NS merger) scenario.} The second r-process nucleosynthesis occurs when the two NS merge, as in the kilonova r-process (e.g., \citealt{Korobkinetal2012}). 

\textit{\#BH The collapsar scenario.} If some core mass stays bound, then the BH remnant of the NS-NS merger accretes mass and might launch jets. The conditions inside the jets might allow r-process nucleosynthesis as in the collapsar r-process scenario that \cite{Siegeletal2019} proposed. 

In the vast-majority of proposed r-process nucleosynthesis events these three r-process sites appear as separate violent events, each that can lead to nucleosynthesis of (some) r-process isotopes. \cite{GrichenerSoker2019b} thoroughly compare these three separate r-process nucleosynthesis scenarios with each other and to observations (see also \citealt{Tarumietal2021}), and I will not repeat the details of that study, but rather only discuss the implications of the scenario I study here. I also comment that I accept the notion that likely more than one type of r-process site contributes to r-process nucleosynthesis (e.g., \citealt{Coteetal2019, GrichenerSoker2019b}). 

The proposed NS-NS merger in CEJSN scenario might have two implications for r-process nucleosynthesis in the early Universe. (1) The proposed scenario substantially shortens the time to NS-NS merger in the kilonova scenario, and (2) the scenario might be a site of a large r-process nucleosynthesised mass that comes from three different r-process sites. 

One of the drawbacks of the kilonova r-process site is that it starts to take place at a relatively long time after the NS-NS formation, i.e., the gravitational waves timescale. Studies show the need for some r-process nucleosynthesis events in the early Galaxy (e.g., \citealt{BeniaminiHotokezaka2020}) and more generally in the early Universe.    \cite{Bonettietal2019}, for example, argue that because of this long delay to NS-NS merger the kilonova scenario cannot account for the r-process abundance of old stellar populations in ultrafaint dwarf galaxies. On the other hand, some studies find solutions to the problems of the kilonova scenario  (e.g., \citealt{Beniaminietal2016, Safarzadehetal2019a, BeniaminiPiran2019, Tarumietal2021}). 

\cite{Safarzadehetal2019a} argue that in one of the model they studied a fraction of $\simeq 0.2$ of NS-NS binaries can merge within 1Myr, and that this solves the problem of long delay time in the kilonova r-process scenario. The ratio of early NS-NS merger events to total NS-NS merger events that I found in equation (\ref{eq:EarlyToNSNS}) is lower by a factor of about 7 than the required early merger rate according to \cite{Safarzadehetal2019a}. This gives a small contribution to the required early merger, but a non-negligible one. I therefore encourage further study of the early merger in the NS-NS merger in CEJSN scenario.  
  
I note two other processes that have been mentioned in the literature in relation to the evolutionary routes that I study here. \cite{Thoneetal2011}, for example, suggest that the jets that a single NS launches in a CEJSN might lead to a long GRB. The same might occur with jets from two NSs in the present study. Such GRBs would not be standard GRBs because they are expected to be very long, including a long-lasting X-ray emission, and they will present hydrogen lines. They might possibly be similar to the gamma-ray and X-ray transient Swift 1644+57 (e.g., \citealt{QuataertKasen2012}). 

\cite{GlanzPerets2021} mentioned that the NS/BH-NS/BG merger process in a CEJSN leads to gravitational-waves emission with unique signatures due to the role of the dynamical friction in reducing the orbit. They suggest that these signatures are somewhat similar to those in the merger of a single NS/BH with the core of a RSG during a CEE, a process that \cite{Ginatetal2020} have studied. I emphasise the long and luminous event that comes along with this gravitational waves source.  

\section{Other triple-star CEJSN cases} 
\label{sec:OtherCases}

I list the NS-NS merger in CEJSN scenario that I study here and several other (but not all) possible triple-star CEJSN events in Table \ref{Table:Scenarios}. The listed properties of the triple star systems do not by themselves determine the light curve and properties of the CEJSN event. The properties of the pre-CEE mass ejection and the RSG envelope mass and angular momentum also influence the light curve and other properties of the observed CEJSN event \citep{SokeretalGG2019, Schroderetal2020, Soker20212CEJSNe}, as well as the possibility of a GRB (e.g., \citealt{Thoneetal2011}). A GRB event might accompany all scenarios that I list here, but I do not write this in the table. In addition to the `Standard CEJSN' in binary systems, \cite{SokeretalGG2019} consider cases of binary star evolution where the RSG envelope is of low mass and highly oblate (Polar CEJSN), of very low mass $<1 M_\odot$ (stripped-envelope CEJSN), the RSG is very massive and the explosion converts a large fraction of the energy to radiation (prolonged CEJSN), and the CEJSN r-process scenario. Namely, each of the cases that I list in Table \ref{Table:Scenarios} can be further classified into two or more sub-cases according to the properties of the RSG envelope and CSM at explosion. I will not discuss these sub-cases in the present study. 
I base some of the outcomes that I list in the table on earlier studies of triple star CEE of other types of triple star systems (e.g., \citealt{SabachSoker2015, Hilleletal2017, ComerfordIzzard2020, GlanzPerets2021, Hilleletal2017}). 
\begin{table*}
\caption{Triple-star CEJSN cases}
\centering
\footnotesize
\hskip -1.90 cm
\begin{tabular}{| l | c | c | c | c | c |c|}
\hline
Name        & Tight         & Wider       & Main jet-powering & Remnant      & Some possible           & Rate to \\ 
            & binary        & star        & events            &              & outcomes                & CCSNe\\
\hline 
Double      &NS     +       & RSG         & 1) NS inside MS   & A BH or      & Super-energetic SN      & $\approx 3 \times 10^{-6}$\\ 
CEJSN       & intermediate  &             & 2) NS inside core & a massive NS & with a strong precursor & $- 3 \times 10^{-5}$\\ 
\citep{Soker20212CEJSNe}& -mass MS &       & [or CCSN]         & [or NS+NS]   & to the main peak;      & \\ 
            &               &             &                   &              & r-process elements.     & \\ 
\hline 
micro-TDE   &  BH   +       & RSG         & 1) BH-accreting   & A BH         & Super-energetic SN      & $\approx 3 \times 10^{-6}$\\ 
CEJSN       & intermediate  &             &  destroyed MS     & [or BH+NS    & with a strong precursor & $- 3 \times 10^{-5}$\\ 
\citep{Soker20212CEJSNe}& to low-mass   & & 2) BH accreting   &  or BH +BH]  & to the main peak;       & \\
            &               &  MS         & core [or CCSN]    &              & $10^{15} \eV$ neutrinos &   \\ 
\hline   
NS-NS merger& NS + NS       & RSG         & 1) NS-NS merger   &  A BH        & Super-energetic SN;     & $\la 10^{-6}$ \\ 
in CEJSN    &               &             & 2) NSs/BH in core & [or BH + NS  & Extreme r-process site; & $- 2 \times 10^{-5}$\\ 
(This study)&               &             &    or  CCSN.      &  or BH +BH]  & $10^{15} \eV$ neutrinos &  \\
\hline   
BH-NS merger& BH + NS       & RSG         & 1) BH-NS merger   &  A BH        & Super-energetic SN;     & ?$\la 2 \times 10^{-7}$ \\ 
in CEJSN    &               & $\ga 30 M_\odot$ & 2) BH inside core & [or BH + NS  & r-process site     & $- 4 \times 10^{-6}$?\\ 
(Thsi study)&               &             &  [or CCSN]     &  or BH +BH]  & $10^{15} \eV$ neutrinos &    \\
\hline   
BH-BH merger& BH + BH       &    RSG      & 1) BH-BH merger   &  A BH        & Super-energetic SN;     & ?$\la 2 \times 10^{-7}$\\ 
in CEJSN    &               &$\ga 30 M_\odot$& 2) BH inside core& [or BH + NS& r-process site          & $- 4 \times 10^{-6}$?\\ 
(This study)&               &             &  [or CCSN]     & or BH +BH]   & $10^{15} \eV$ neutrinos &    \\
\hline   
`Name'      & Any of the    & RSG         & Only one NS/BH    & NS+MS or BH+MS & Regular CEJSN         & \\ 
ejection    & above         &             & inside core       & or NS/BH + NS/BH &  with an ejected    & \\ 
CEJSN       & binaries      &             & [or CCSN]         & [NS/BH+NS/BH+MS] &  NS/BH or a MS      & \\ 
 \hline   
Eccentric   & NS/BH         & Any that    &NS enters and exists& Triple with a hot NS& A CEJSN-impostor & \\ 
CEJSN       & + RSG         & perturbs     & the RGB envelope  &  and an RGB with    & event that might &   \\ 
impostor    &               & orbits      &                   &  reduced mass       & repeat [G19; Sc21]& \\ 
\hline   
\end{tabular}
\footnotesize
\begin{tablenotes}
Some CEJSN cases in triple-star systems. The tight binary system (inner binary) is the system of the short orbital period at the onset of the CEE, while the wider star is the one that engulfs the tight binary system, or in the last row it is the star that perturbed the orbit of the tight binary system that enters a binary (intermittent) CEE. 
In all cases the NS and/or BH accrete mass from the RSG envelope and launches jets that add to the powering of the CEJSN; I do not list this accretion phase in the table, but rather only higher accretion rates phases from a MS or from the core of the RSG.  Whenever a NS accretes from an oxygen-rich core r-process nucleosynthesis might take place.  
Inside square parentheses are the outcomes in case the NS or BH do not enter the core of the RSG and the core explodes as a CCSN (the second or the third CCSN in the triple star system). 
In the last column I list estimates of the event rate to that of the CCSN rate. Question marks in the last column indicates speculative and highly uncertain rate estimates.  
References: G19: \cite{Gilkisetal2019}; S21: \cite{Schreieretal2021}.
Abbreviation: BH: black hole; CEE: common envelope evolution; MS: main sequence (star); NS: neutron star; RSG: Red supergiant; TDE: tidal disruption event. 
\end{tablenotes}
\label{Table:Scenarios}
\end{table*}

I list outcomes of cases where the NS/BHs (singles or binaries) manage to eject the envelope and do not penetrate or destroys the core inside square parentheses. These cases are also termed CEJSN impostors, although I do not always use this term in this study. 
   
\textit{Double CEJSN.} In this scenario that I explored in \cite{Soker20212CEJSNe} a NS-MS tight binary system enters the RSG envelope, followed by a CEE of the NS inside the MS star. The NS launches jets as it accretes mass from the dense MS star and launches energetic jets. Hence the name double CEJSN. The light curve of this event might have two very bright and long peaks. The table lists some more properties and some possible outcomes (for more details see \citealt{Soker20212CEJSNe}).  

\textit{Micro-TDE CEJSN.} A similar evolution to that of the double CEJSN might take place if instead of a NS the tight binary system is of a BH and a MS star when it enters the envelope of the RSG \citep{Soker20212CEJSNe}. The BH is several times the mass of the NS, and in many cases it will disrupt the MS star before it enters its intact envelope. The MS material forms a massive accretion disk around the BH, and the BH launches very energetic relativistic jets. The process by which a stellar BH disrupt a star and accretes its mass is termed a micro-tidal disruption event (micro-TDE; \citealt{Peretsetal2016}). 
If the MS is very massive, the evolution might be through a BH-MS CEE, like for the NS. As well, if the MS is of low mass in the double CEJSN scenario, then the NS might disrupt the MS star before the systems enters a CEE. However, in most relevant cases a NS-MS tight binary system will enter a CEE and a BH-MS system  will suffer a micro-TDE.  Overall, the light curve of a micro-TDE CEJSN might be similar to that of double CEJSN event, but much more energetic and with relativistic jets. 
One possible outcome of a BH accreting from the envelope of a RSG is the production of $\approx 10^{15} \eV$ neutrinos \citep{GrichenerSoker2021}.  

\textit{NS-NS merger in CEJSN.} This is the scenario that I studied in previous sections. I presented four evolutionary routes of the case where a NS-NS tight binary system enters an RSG envelope in Fig. \ref{fig:SchematicScenario}, where in one of them the two NSs do not merger (phase P2-A). There are other evolutionary routes not shown in the figure where the two NSs do not merge, one in which one of the two NS is ejected from the system (see below) and one in which the two NSs manage to eject the envelope before they enter the core. The outcome in the last case is either a system (bound or unbound) of three NSs or of a BH with two NSs. 

The basic process of the NS-NS merger in CEJSN scenario is that the friction within the RSG envelope and some mass accretion might bring the NSs close enough to merge, resulting in gravitational waves and the launching of energetic jets. Here I only add that if the two NSs merge in the envelope then the light curve might have two bright peaks, one peak after the NS-NS merge and the second after the merger remnant enters the core, or, alternatively if the merger remnant strips the RSG envelope, after the core explodes as a type Ib or Ic CCSN.  

\textit{BH-NS merger in CEJSN} and \textit{BH-BH merger in CEJSN}. These are similar to the NS-NS in CEJSN scenario, but one or two of the NSs are replaced by BHs. For the BH to merge inside the envelope with a NS or another BH, the RSG should be very massive, $M \ga 30 M_\odot$ \citep{Schroderetal2020}. This would lead to a very long and extremely energetic CEJSN, but an extremely rare one. I take each of these two cases to be less frequent than the NS-NS merger cases as they require much more massive RSG stars. However, at this time I can only speculate on the relative number of events. I take the ratio of the number if RSG stars with initial mass of $M_{\rm ZAMS} \ga 10 M_\odot$, which can engulf NS-NS tight binary system, to the number of more massive stars of $M_{\rm ZAMS} \ga 30 M_\odot$, which might be able engulf binaries with a BH or two.  I take this masses ratio as the minimum BH mass is about three times that of a NS. I therefore estimate the number of each of the scenarios with BH-NS and BH-BH binaries that enter RSG envelopes to be $\approx 0.2$ times that of the NS-NS in a CEJSN scenario. This estimate does not take onto account mass transfer that can make the RSG much more massive than its initial mass. Because these estimates are more of a speculation, I boarder each value with two question marks.  

\textit{`Name' ejection CEJSN.} Here the `Name' is any of the names above, but the word ejection replaces the word `merger'. In this type of scenarios  the tight binary system that enters the RSG breaks up inside the CEE (e.g., \citealt{SabachSoker2015, GlanzPerets2021}), and at most one NS/BH enters the core. The other component of the tight binary system (a MS star, or a NS, or a BH) is ejected from the envelope. It might stay bound on a wide orbit, it might stay bound and fall back to the RSG envelope, or it might escape the system altogether. The closer bound NS/BH might merge with the core or eject the envelope leaving a core-NS/BH close system. In the later case the core will explode as a CCSN, the third one in the triple system (unless there is a MS star). 
As in all cases, when the tight binary system is of NS/BH-NS/BH, the two stars can launch jets together, leading to a very complicated mass loss morphologies and complicated light curves. The calculation of the rates of the ejection evnets is beyond the scope of this study.  

\textit{Eccentric CEJSN impostor.} This scenario differs from the other cases in that the inner binary is composed of the NS/BH and the RSG star. The tertiary star serves only to perturb the NS/BH such that it acquires a highly eccentric orbit and enters the RSG envelope and exists from it. While inside the envelope it accretes mass via an accretion disk and launches jets that power a bright event, called CEJSN impostor. \cite{Gilkisetal2019} proposed this scenario and \cite{Schreieretal2021} simulate the interaction of the jets with the envelope. The process might repeat itself, and the system might later enter a continues CEE towards a CEJSN (if the NS/BH enters the core of the RSG). 
  
In all these cases, jets might also be a source of gamma ray, leading to a not standard GRB as in binary systems (e.g., \citealt{FryerWoosley1998, Thoneetal2011, ZhangFryer2001}). 

The cases that I listed above (Table \ref{Table:Scenarios}) together with the different possible properties of the RSG envelope and CSM at the onset of the CEJSN \citep{SokeretalGG2019}, show that CEJSNe of triple-star systems can yield a rich variety of peculiar super-energetic SNe. This calls for the attention by future discoverers of peculiar SNe to the CEJSN scenarios, in binary and in triple star systems.

\section{Summary} 
\label{sec:Summary}
 
I proposed the \textit{NS-NS merger in CEJSN} scenario and explored some of its properties and evolutionary routes. The key evolutionary ingredients of this scenario where a tight NS-NS binary system enters a CEE with a RSG star are (1) the  launching of jets by the two NSs and (2) their merger, either in the envelope or in the core of the RSG star. The CEE before merger involves the launching of jets by the two NSs, and after merger by the merger remnant, inside the envelope and possibly inside the core of the RSG star, as the compact objects accrete mass from the envelope and core. These jets by themselves power a CCSN-like transient event that is termed a CEJSN (strictly speaking, if the NS/BH do not enter the core the name is CEJSN impostor). In the NS-NS merger in CEJSN scenario there is in addition the energy that the merger process deposits to the RSG star. Overall, the NS-NS merger in CEJSN events can reach a total energy (kinetic + radiation) of $E_{\rm tot} \ga 10^{52} \erg$. 

The light curves of NS-NS merger in CEJSN events are very complicated due to the NS-NS merger, the jets before and after the merger, the collision of shells from consecutive mass ejection episodes with each other, and from the CCSN explosion of the core in cases where the NS/BH do not get into the core. 

In Fig. \ref{fig:SchematicScenario} I presented four evolutionary channels. One channel (ending in phase P2-A) does not involve NS-NS merger. There are two other evolutionary channels that are not in Fig. \ref{fig:SchematicScenario} and also do not involve merger (section \ref{sec:OtherCases}). In the first of these two channels, the two NSs do not merger and do not enter the core leaving an intact core that explodes to form an NS or a BH. In the second, the NS-NS ejection CEJSN event, one of the two NSs is ejected from the CEE (it might stay bound or it might become unbound from the system), and the RSG and the closer NS evolves as a binary CEJSN (or impostor).

In section \ref{sec:rate} I crudely estimated the rate of NS-NS merger in CEJSN events relative to the rate of CCSNe 
$f_{\rm EMe/CCSN}  \la 10^{-6} - 2 \times 10^{-5}$ (equation \ref{eq:eq:EarlyToCCSN}).

In section \ref{sec:r-process} I discussed the implications to  r-process nucleosynthesis in the early Universe (and Galaxy). Three of the evolutionary routes that I presented in Fig. \ref{fig:SchematicScenario} involve NS-NS merger. The NS-NS merger is a site of r-process nucleosynthesis, the kilonova r-process scenario. The launching of jets by a NS in a post-helium burning core is another r-process site, the CEJSN r-process scenario. The launching of jets by a BH as it accretes at a high rate from a massive accretion disk is the collapsar r-process scenario. The extremest channel I studied here is the one depicted in the left column of Fig. \ref{fig:SchematicScenario} because it might involve these three r-process sites. 

CEJSNe lead to early NS-NS merger events compared with the merger timescale due to gravitational waves. This might ease the problem of the kilonova r-process scenario to explain r-process isotopes in the early Universe. It seems that although the NS-NS merger in CEJSN scenario plays a role in the r-process nucleosynthesis in the early Universe, it cannot solve this problem by its own. Future studies of r-process nucleosynthesis in the young Galaxy, and in the early Universe in general, should include this scenario, as well as the CEJSN r-process scenario in binary stars.

The observational signatures of NS-NS merger in CEJSN events  are the complicated, long, and very luminous light curve, the peculiarity of newly synthesised (radioactive) r-process isotopes, the emission of gravitational waves that accompany the event, and a BH remnant.  

In section \ref{sec:OtherCases} I summarise some triple-star CEJSN events (Table \ref{Table:Scenarios}) that place the NS-NS merger in CEJSN scenario in a bigger picture.
The NS-NS merger in CEJSN and each of the other similar scenarios that I list in Table \ref{Table:Scenarios} is very rare. However, it is important to study them in light of the expected very large number of luminous transient events that existing an upcoming sky surveys will detect, e.g., the Large Synoptic Survey Telescope (LSST; \citealt{Ivezicetal2019}), the Zwicky Transient Facility (ZTF; \citealt{Bellmetal2019}), the All-Sky Automated Survey for Supernovae (ASAS-SN; \citealt{Kochaneketal2017PASP}), and the Southern Hemisphere Variability Survey (LSQ; \citealt{Baltayetal2013}).  
These are expected together to observe $\approx 10^4$ event per year. Because the typical CEJSN event is more luminous that a typically CCSN, I expect the detected rate of CEJSN events to be larger than the ratios that I list in the last column of Table \ref{Table:Scenarios}. Crudely, I expect that in the next decade tens to hundreds of peculiar and puzzling luminous transients could be classified to belong to one of the triple-star CEJSN events.

\section*{Acknowledgments}

I thank Aldana Grichener for helpful discussions and comments, and an anonymous referee for useful suggestions. 
This research was supported by the Amnon Pazy Research Foundation.

\textbf{Data availability}
The data underlying this article will be shared on reasonable request to the corresponding author.  


\label{lastpage}
\end{document}